\documentclass[11pt,a4paper]{article}
\usepackage{fullpage}
\usepackage{graphicx}
\usepackage{setspace}
\usepackage{natbib}
\usepackage{graphicx}
\usepackage{amsmath}
\usepackage{amsfonts}
\usepackage{amssymb}
\usepackage{color}
\usepackage[english]{babel}
\usepackage{multirow}
\usepackage{booktabs}
\usepackage{pdflscape}
\usepackage{threeparttable}
\usepackage{tabularx}
\usepackage{url}
\usepackage{pgffor}
\usepackage{ifthen}
\usepackage{float}
\usepackage{svg} 
\usepackage{dcolumn} 
\makeatletter 
\usepackage{bm}
\newcommand*{\rom}[1]{\expandafter\@slowromancap\romannumeral #1@}
\makeatother

\renewcommand{\baselinestretch}{1.5}

\title{A tale of two tails: 130 years of growth-at-risk\footnote{We would like to thank participants of the 11$^{th}$ European Seminar on Bayesian Econometrics and the Annual Meeting of the Austrian Economic Association for comments and suggestions. Huber gratefully acknowledge financial support from the Austrian Science Fund (FWF, grant no. ZK 35) and the Oesterreichische Nationalbank (OeNB, Anniversary Fund, project no. 18304). The views expressed in this paper are those of the authors and do not necessarily represent those of the Liechtenstein Financial Market Authority.}} 
\author{}
\date{\today}

\author{Martin Gächter\footnote{Liechtenstein Financial Market Authority, Liechtenstein; University of Innsbruck, Austria.}
\hspace{1.5cm}
Elias Hasler\footnote{Corresponding author: Liechtenstein Financial Market Authority, Liechtenstein; University of Innsbruck, Austria; e-mail: Elias.Hasler@fma-li.li.}
\hspace{1.5cm}
Florian Huber\footnote{University of Salzburg, Austria}
}

\begin{document}

\maketitle

\renewcommand{\baselinestretch}{1.5}

\maketitle

\begin{abstract}
\noindent We extend the existing growth-at-risk (GaR) literature by examining a long time period of 130 years in a time-varying parameter regression model. We identify several important insights for policymakers. First, both the level as well as the determinants of GaR vary significantly over time. Second, the stability of upside risks to GDP growth reported in earlier research is specific to the period known as the Great Moderation, with the distribution of risks being more balanced before the 1970s. Third, the distribution of GDP growth has significantly narrowed since the end of the Bretton Woods system.  Fourth, financial stress is always linked to higher downside risks, but it does not affect upside risks. Finally, other risk indicators, such as credit growth and house prices, not only drive downside risks, but also contribute to increased upside risks during boom periods. In this context, the paper also adds to the financial cycle literature by completing the picture of drivers (and risks) for both booms and recessions over time.

\bigskip


\bigskip
\bigskip
\end{abstract}
\newpage

\section{Introduction}
The empirical growth-at-risk (GaR) concept introduced by \cite{adrian2019vulnerable} suggests that deteriorating financial conditions are associated with increased downside risks to economic growth. While standard forecasts focus on the expected value of future GDP growth, the GaR approach places a particular emphasis on the probability and magnitude of potentially adverse outcomes. Similar to the value-at-risk concept in finance, the GaR of an economy for a given time horizon is defined as a specific low quantile of the distribution of the projected GDP growth rate for the respective horizon. In this context, \cite{adrian2019vulnerable} show that the left tail of the distribution of (projected) GDP growth is less stable and more affected by financial conditions than the upper quantiles of the distribution. Against this background, the GaR concept is a useful and intuitive policy tool to identify and quantify systemic risk and has therefore gained traction among policy-makers in recent years.

In the last few years, the GaR idea has been extended in various directions, e.g. adding various risk indicators from the financial cycle literature \citep[see, for instance, ][]{aikman2019credit} or by examining the term structure of GaR \citep{adrian2018}. In this context, the GaR framework is used as a composite indicator for systemic risk at the country level and can, therefore, also be taken as an indicator when to activate various policy measures. Consequently, recent research has also taken into account the impact of various policy instruments on GaR \citep[e.g.][]{galan2020}. Remarkably, the entire strand of literature has relied entirely on data samples back to the 1970s without taking into account earlier developments in the last century.\footnote{One notable exception that relies on copula models is \cite{coe2020financial}. As opposed to this paper, they focus on the role of financial conditions for tail risks to economic growth.} This is insofar surprising, as the main objective of macroprudential policy is the prevention of financial crises (or alternatively the reduction of the costs of such crises if they occur). Since financial crises appear infrequently, as also shown by the widely cited financial cycle literature \citep{schularick2012credit,jorda2017}, a long time series of the underlying drivers is crucial to capture the tail risks of the variables of interest. 

Using long time series, however, raises additional econometric issues. For instance, these time series might be subject to structural breaks in the conditional mean. Such a behaviour might reflect changes in key relationships between the conditional distribution of output growth or long-run unconditional means. In addition, the volatility of shocks is often found to be time-varying \citep[for predictive evidence of this claim, see, e.g.,][]{clark2011real}. Standard quantile regressions (QRs) have difficulties matching both features of historical data. In this paper, we build on recent papers that show that simple heteroskedastic models perform well (or even better) than QRs \citep{carriero2020capturing,brownlees2021} and  propose using a time-varying parameter stochastic volatility regression model (TVP-SV) to capture changes in the relevance of different potential drivers of upside and downside risks to output growth. We use this framework to analyze an extended data set covering 130 years of macroeconomic data and show that it works well for predicting downside risks to GDP. After providing evidence that such a model is competitive to standard QRs, we back out the contributions of individual variables over time using state-of-the-art techniques from statistics and machine learning \citep[see, e.g.,][]{crawford2019variable,woody2021model}.\footnote{\citet{clark2022forecasting} use a similar approach to summarize the effect of different predictors on inflation within a nonparametric model.} Thus, by applying novel methods to historical data, we are able to draw important policy implications for today's policy-makers. 

Our findings allow us to draw a whole range of relevant policy implications and to put  previous findings into a historical context. First, we show that the stability of upside risks to GDP growth shown in previous research is time-specific for the time period since the start of the Great Moderation, with a more or less symmetric distribution of risks up to the 1970s. Second, both upside and downside risks to GDP have decreased substantially over time, remaining at relatively low levels since the beginning of the Great Moderation. Third, we show that financial stress is always associated with higher downside risks, although the effect varies in magnitude over time, while financial stress does not affect upside risks. Fourth, while the effect of credit growth and house price growth varies over time, the similarities of the effect of credit growth during the Great Depression and the Great Financial Crisis are remarkable. Fifth, we find that the large negative impact of house prices on growth risks during the Global Financial Crisis was unprecedented. Finally, our findings suggest that credit growth and house prices do not only drive downside risks but also increase upside risks, i.e. they lead to a wider distribution of (expected) GDP growth. Against this background, our findings also add to the financial cycle literature by looking at the whole distribution of (expected) GDP, thus completing the picture of drivers (and risks) for booms and recessions over time. In this context, a better understanding of how individual variables and risk indicators influence both upside and downside risks to GDP is crucial for the calibration and timing of macroprudential (and also monetary) policy measures. 

The remainder of the paper is structured in the following way. The next section fleshes out the dataset and econometric methods adopted. Section \ref{sec: results} includes the empirical findings and consists of an out-of-sample tail forecasting exercise, provides quantitative evidence on the predictive distributions of GDP growth and then proceeds by discussing the key drivers of GaR. Section \ref{sec: discussion} puts our empirical findings in context and draws relevant policy conclusions while the final section concludes the paper. 

\section{Data and Methods}
\subsection{Data}
Our analysis is based on a newly constructed data set stretching from 1893Q1 to 2016Q4. Using three different sources, our data includes annualised real GDP growth, a financial stress indicator, the 3-year average growth rate of the credit-to-GDP ratio, and the 3-year average growth rate of real house prices. The real GDP time series is taken from \citet{d2012} and updated using FRED data. 
The dependent variable is constructed by using the logarithm of real GDP, $Y_t$, and converting it into the annualised growth rate $h$ periods ahead, $y_{t+h} = \frac{Y_{t+h} - Y_t}{h/4} $. 

In line with previous literature, we include a measure of financial stress as an explanatory variable. However, unlike other financial stress indicators, which are typically based on financial data and start in the 1970s we use a historical newspaper-based financial stress indicator built by \citet{puttmann2018}. As \citet{puttmann2018} notes, the index exhibits a long-run trend; hence we detrend the time series using a slow-moving Hodrick-Prescott filter with a $\lambda$ of $5 \times 10^6$.

While financial stress measures are highly relevant for short-term GaR estimations -- at least since the 1970s -- credit-to-GDP growth and house prices are frequently used as a signal for medium-term financial imbalances \citep{aikman2019credit, galan2020}. Unfortunately, consistent quarterly data for credit (loans to the non-financial private sector) and house prices are not available for such a long time horizon. Therefore, we use data from \cite{jorda2017} and convert the annual data into quarterly by using the quadratic spline of \cite{forsythe1977}. Subsequently, annualized three-year averages of the log differences of the credit-to-GDP ratio as a measure of credit growth and the three-year average of the log differences of real house prices as a measure of house price growth are used. 

\subsection{Growth at risk through  the lens of TVP regressions}
Downside risks are typically analyzed through QRs \citep[see][]{adrian2019vulnerable}. However, one key shortcoming of QRs is that they are not able to capture structural breaks in the regression coefficients, a feature that is crucial given the length of the time series we analyze. A natural way of capturing time-variation in the parameter is through time-varying parameter (TVP) regression models. These models assume that the coefficients evolve smoothly over time and thus capture changes in transmission mechanisms but, conditional on appropriate modeling assumptions, also allow for rapid movements in the underlying parameters. To capture changing volatilities of the shocks, we also allow for heteroskedasticity in the regression model through a standard stochastic volatility specification. Our simple, yet flexible model enables us to capture differences in the relations between the determinants of GaR but also allows for situations where large, unobserved shocks are the main drivers of tail risks.

In its general form, we consider predictive equations with drifting parameters that take the following form:
\begin{equation}
    y_{t+h} = \bm \beta'_{t+h} \bm x_t + \varepsilon_{t+h}, \quad \varepsilon_{t+h} \sim \mathcal{N}(0, \sigma_{t+h}^2) \label{eq: tvpsv}
\end{equation}
where $\bm \beta_{t+h}$ is a vector of TVPs which link $y_{t+h}$ to our set of $K$ macro-financial covariates in $\bm x_t$. We follow much of the literature \citep[see e.g.][]{primiceri2005, cogley2005drifts} and assume that these evolve according to random walk processes. Moreover, the logarithm of the  error variance is assumed to follow  an AR(1) process. These assumptions give rise to a system of state evolution equations:
\begin{align*}
    \bm \beta_t &= \bm \beta_{t-1} + \bm \eta_t, \quad \bm \eta_t \sim \mathcal{N}(\bm 0_K, \bm V_\beta),\\
    \log \sigma_t^2 &= \mu_\sigma + \rho_\sigma (\log \sigma_{t-1}^2 - \mu_\sigma) + w_t, \quad w_t \sim \mathcal{N}(0, \vartheta^2),\\
    \log \sigma_0^2 &\sim \mathcal{N}\left(\mu_\sigma, \frac{\vartheta^2}{1-\rho_\sigma^2}\right)
\end{align*}
where $\bm V_\beta = \text{diag}(v^2_1, \dots, v^2_K)$ is a diagonal matrix with variances $v_j^2$. These variances control the amount of time-variation in the regression coefficients. If $v_j^2 = 0$, $\beta_{jt}$, the $j^{th}$ element of $\bm \beta_t$, would be constant over time since $\beta_{jt} = \beta_{jt-1}$ for all $t$. Hence, the corresponding effect of $x_{jt}$ on $y_{t+h}$ is time-invariant. By contrast, setting $v_j^2$ to a large value implies substantial variation in the corresponding coefficient, giving rise to overfitting concerns. 

The coefficients associated with the law of motion of the log-volatilities are the long-run unconditional mean $\mu_\sigma$, the persistence parameter $\rho_\sigma$ and the innovation variance $\vartheta^2$. If $\rho_\sigma$ is close to one, the corresponding volatility estimate will be smooth. The variance parameter $\vartheta^2$ controls the amount of time variation in the error variances.

Deciding on whether we need time variation in the parameters is a nonstandard statistical problem and the Bayesian literature offers several solutions. In this paper, we use shrinkage priors to decide on whether coefficients are constant or time-varying. In particular, we use the tripple Gamma shrinkage prior \citep{cadonna2020triple} as implemented in the \texttt{R} package \texttt{shrinkTVP} \citep{knaus2021shrinkage}.  More details on the priors and the posterior simulator are provided in Appendix A.

In this paper, we will take a predictive stance and consider the predictive distribution of the model in (\ref{eq: tvpsv}). The predictive density is given by:
\begin{equation}
    p(y_{T+h}|Data_{1:t}) = \int p(\bm y_{T+h}|\bm \Xi, Data_{1:T}) p(\bm \Xi|Data_{1:T}) d\bm \Xi, \label{eq: pred_dens}
\end{equation}
where $Data_{1:T}$ denotes the available information up to  time $T$, $\bm \Xi$ is a generic object that collects coefficients and latent states.  The predictive density is not available in closed form and obtained through simulation-based techniques using the output from the MCMC sampler.

Notice that $p(\bm y_{T+h}|\bm \Xi, Data_{1:T})$ is Gaussian:
\begin{equation*}
    y_{T+h}|\bm Xi, Data_{1:T} \sim \mathcal{N}(\bm \beta'_{T+h} \bm x_T, \sigma^2_{T+h}),
\end{equation*}
and thus symmetric.  However, once we integrate over $\bm \Xi$ the corresponding predictive distribution takes a non-standard form and accommodates features such as skewness, heavy tails and downside asymmetries as reported in, e.g., \cite{adrian2019vulnerable}. In addition, the fact that the parameters vary over time allows us to investigate whether different elements in $\bm x_t$ vary in importance for explaining growth at risk over time.

\section{Empirical findings}\label{sec: results}
\subsection{Model evaluation and features of the predictive densities}
In a first step, we evaluate the out-of-sample forecasting performance of different TVP and quantile regression models. For both the quantile regression and the TVP model, we estimate two specifications:  \textit{(i)} the baseline model including a constant, the financial stress indicator, and lagged GDP growth as predictors (we will use the abbreviation QR and TVP to refer to these models), and \textit{(ii)} the extended model which additionally includes credit growth and house price growth as regressors (hereafter referred to as QR+ and TVP+).

We evaluate the accuracy of the predicted downside risks with the help of quantile scores \citep[see e.g.][]{giacomini2005, brownlees2021}. Our focus is on the performance in the left tail. Hence, we consider the quantile score at the 5 percent quantile.

Table \ref{table:QS} reports the quantile scores relative to the QR model for the pre and post World War \rom{2} (WW\rom{2}) subsample and the forecasting horizons $h=1$ and $h=4$.\footnote{We need to split the sample because the quantile regression models can not adequately handle the large drop in GDP volatility post WW\rom{2}. Furthermore, \cite{amir2016} show that correlations between variables, forecasts and other statistics change considerably at certain points in time when using long time series, making an a priori choice of subsamples hard to defend. Moreover, \citet{gachter2022} show that the average downside risks and the magnitude of effects of financial risk indicators depend on the structural characteristics of a country. These country characteristics change over time and therefore should be taken into account when not using a TVP approach.}  In general, we find that TVP models perform particularly well during the post-WW\rom{2} period, especially at the one-step ahead horizon. For both periods and horizons considered, the TVP models produce quantile scores that are superior to those obtained from the extended quantile regression. However, it is worth noting that the small QR model produces slightly more precise tail forecasts during the pre-WW\rom{2} period.

Once we focus on four quarter ahead GDP growth forecasts, this  pattern becomes slightly less pronounced. In this case, both quantile regressions (QR and QR+) improve upon the TVP models when the period prior to WW\rom{2} is considered. When we focus on post-WW\rom{2} data, the TVP regressions are again outperforming the simple quantile regression model by appreciable margins. The reason for this rather weak performance in the pre-WW\rom{2} period is driven by the fact that the QRs produce predictive densities with  wide credible intervals. This helps in periods characterized by sharp breaks in GDP growth but harms forecasting accuracy in tranquil times. Since the pre-WW\rom{2} features several crises, a model which produces wide forecast intervals during that specific points in time yields favorable overall tail forecasts. 

\begin{table}[!t] \centering \footnotesize
  \caption{Out-of-sample model evaluation} 
  \label{table:QS} 
\begin{tabular}{@{\extracolsep{3pt}}lD{.}{.}{-1} D{.}{.}{-1} D{.}{.}{-1} D{.}{.}{-1}
D{.}{.}{-1}D{.}{.}{1}D{.}{.}{-1}} 
\\[-1.8ex]\hline 
\hline \\[-1.8ex] 
 & \multicolumn{3}{c}{Pre WW\rom{2}} &  \multicolumn{3}{c}{Post WW\rom{2}}\\ 
\cline{2-4} \cline{5-7} \\
\multicolumn{1}{c}{Horizon} &\multicolumn{1}{c}{TVP} &\multicolumn{1}{c}{TVP+} &\multicolumn{1}{c}{QR+} &\multicolumn{1}{c}{TVP} &\multicolumn{1}{c}{TVP+} &\multicolumn{1}{c}{QR+} \\ 
\hline \\[-1.8ex] 
$h=1$ & 1.063 & 1.074  & 1.229 &  0.856 & 0.853 & 1.010\\
$h=4$ & 1.887 & 1.715  & 1.223  & 0.937 & 0.886 & 1.117 \\
\hline 
\hline \\[-1.8ex] 
\end{tabular} 
\begin{minipage}{1\textwidth} 
{\footnotesize  \textit{Note:} This table reports the out-of-sample model evaluation for the forecast horizon of 1 and 4 quarters. The quantile scores are reported relative to the QR model.} 
\end{minipage}
\end{table}

This brief discussion has shown that TVP models can outperform quantile regressions, especially at the one quarter ahead horizon, but also for multi-step ahead forecasts and post-WW\rom{2} data. Since we focus on two variants of the models (i.e. the baseline and the extended versions), we can also analyze whether including credit and house price growth pays off for obtaining more precise tail forecasts.  The results in Table \ref{table:QS} suggest that for one quarter ahead tail forecasts, additional information does not translate into more precise predictions for the pre-WW\rom{2} sample and only slightly improves the predictive fit over the post-WW\rom{2} period. This pattern reverses if our interest is on four step ahead forecasts. In that case, using more information yields more precise forecasts (which are still inferior to the QR benchmark predictions pre-WW\rom{2}) for both hold-out periods considered. This finding is most likely driven by the fact that one quarter ahead predictions are dominated by high frequency shocks which are notoriously difficult to predict whereas for longer-run forecasts, short-lived trends become more important.

In sum, our small forecasting exercise shows that TVP regressions are capable of producing competitive tail forecasts without explicitly focusing on the corresponding quantile under scrutiny. This indicates that the increased flexibility provided by TVP models (i.e. allowing for drifts in $\bm \beta_t$ and changing error variances in $\sigma_t^2$) enables us to adequately model the  distribution of GDP growth over long samples.

Next, we examine the characteristics of the predictive densities. Figure \ref{fig:pred1q} presents the annualized one- and four-quarter ahead GDP growth along with their predicted lower ($5^{th}$ percentile) and upper ($95^{th}$ percentile) bounds. The results for the post-1970s period are consistent with previous research, which shows that lower bounds vary significantly over time while upper bounds are relatively stable \citep[see, for example,][]{adrian2019vulnerable, aikman2019credit}. However, this pattern appears to have emerged only since the start of the Great Moderation period in the 1980s. Prior to that, upper bounds were just as volatile as lower bounds. Table \ref{table:sd} supports this observation by showing the standard deviation of up- and downside risks over four different periods. The table also highlights the significant reduction in overall tail volatility after WW\rom{2} and the even lower variation in the tails since the Great Moderation. This is not surprising, as the Great Depression led to a change in policy-making, with a greater focus on stabilizing the business cycle and reducing volatility.

\begin{figure}[!htbp]
\includegraphics[width = 1\textwidth]{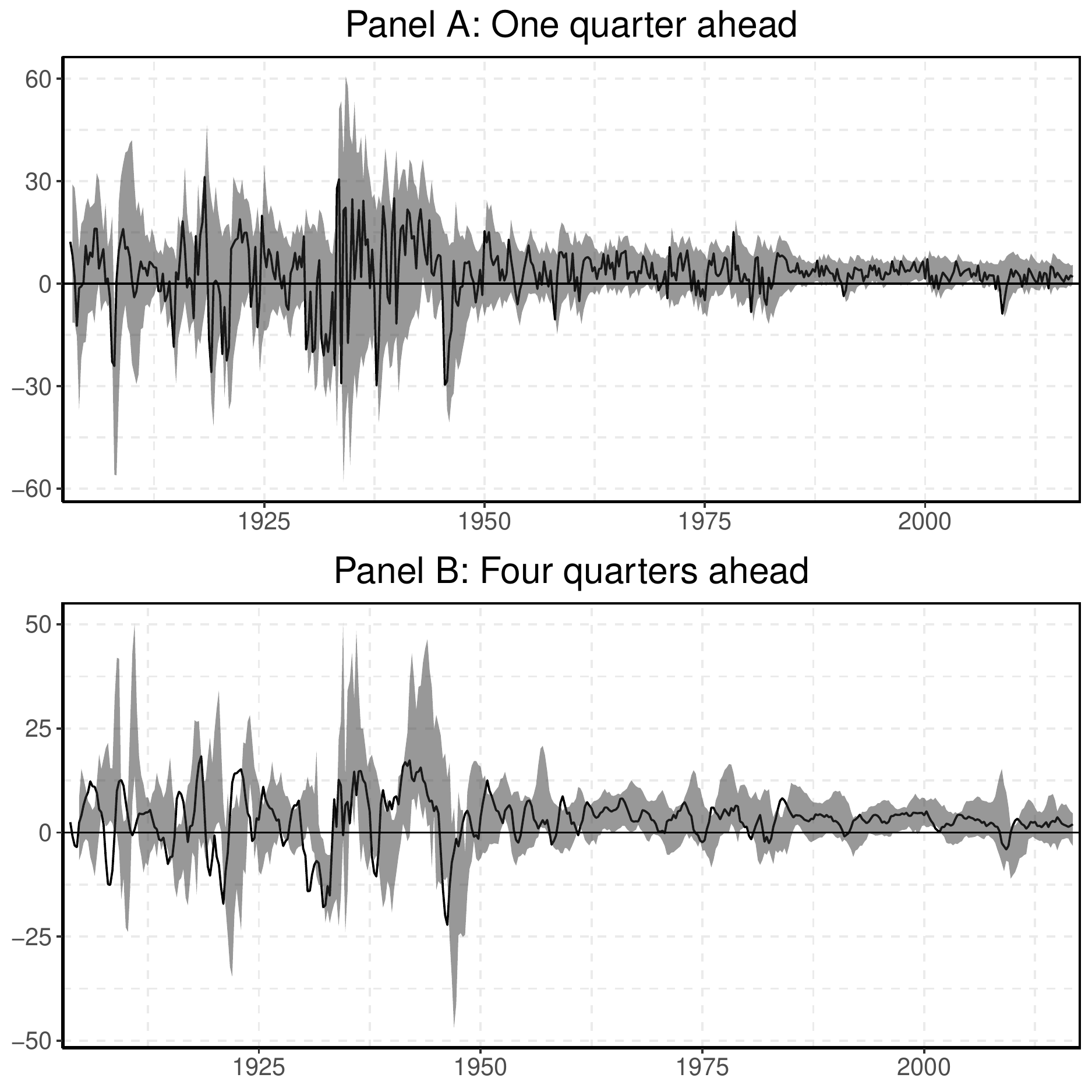}
              \caption{Time series evolution of the predicted tail risks}
              \label{fig:pred1q}
    \medskip 
    \begin{minipage}{1\textwidth} 
		{\footnotesize \textit{Note:} This figure shows the out-of-sample one quarter ahead (Panel A) and four quarters ahead (Panel B) forecast of the $5^{th}$ and $95^{th}$ percentile together with the realised growth rate.\par}
	\end{minipage}
\end{figure}

\begin{table}[!htbp] \centering \footnotesize
  \caption{Standard deviation of up- and downside risks} 
  \label{table:sd} 
\begin{tabular}{@{\extracolsep{3pt}}lD{.}{.}{-1} D{.}{.}{-1} D{.}{.}{-1} D{.}{.}{-1}
D{.}{.}{-1}D{.}{.}{-1}D{.}{.}{-1}D{.}{.}{-1}D{.}{.}{-1}} 
\\[-1.8ex]\hline 
\hline \\[-1.8ex] 
 & \multicolumn{4}{c}{Upside risks} &  \multicolumn{4}{c}{Downside risks}\\ 
\cline{2-5} \cline{6-9} \\
\multicolumn{1}{c}{Horizon} &\multicolumn{1}{c}{pre WW\rom{1}} &\multicolumn{1}{c}{Interwar} &\multicolumn{1}{c}{pre GM} &\multicolumn{1}{c}{since GM} &\multicolumn{1}{c}{pre WW\rom{1}} &\multicolumn{1}{c}{Interwar} &\multicolumn{1}{c}{pre GM} &\multicolumn{1}{c}{since GM} \\ 
\hline \\[-1.8ex] 
$h=1$ & 11.03 & 11.74 & 3.09 & 1.82 & 12.78 & 11.62 & 2.87 & 2.60\\
$h=4$ & 11.96 & 11.66 & 3.62 & 2.22 & 7.47 & 9.29  & 3.47 & 2.63\\
\hline 
\hline \\[-1.8ex] 
\end{tabular} 
\begin{minipage}{1\textwidth} 
{\footnotesize  \textit{Note:} This table reports the standard deviation of the predicted $5^{th}$ and $95^{th}$ percentile. The sample is split into four periods: pre WW\rom{1}, the interwar era, the time before the Great Moderation (pre GM) and since the Great Moderation (since GM). After both world wars two years are left out to not capture any wartime effects in the standard deviations.} 
\end{minipage}
\end{table}

\subsection{Decomposing up- and downside risks}

Next, we focus on the drivers of tail risks over time. Our TVP model with stochastic volatility is able to handle asymmetries in up- and downside risks, which is a necessary feature for this type of analysis. The asymmetries in tail risks imply asymmetries in the unconditional distributions but do not necessarily require asymmetries in conditional predictive distributions \citep{carriero2020capturing}. Hence, only focusing on the time series evolution of the coefficients (see Figure \ref{fig:coef1q} and \ref{fig:coef4q}) does not provide a complete picture of how tail risks occur. To shed light on which indicator drives the predictive quantiles of GDP growth, we rely on linear posterior summaries \citep[see, e.g., ][]{woody2021model}. This is achieved as follows. Based on the $h$-step-ahead predictive distribution of the model (see Eq. (\ref{eq: pred_dens})) we compute a sequence of quantiles $\mathcal{Q}_{t+h, p}$. Each of these estimated quantiles is then used as the dependent variable in the following linear regression model:
\begin{equation*}
    \mathcal{Q}_{t+h, p} = \bm \alpha'_p \bm x_t + e_t,
\end{equation*}
where $\bm \alpha_p$ is a quantile-specific set of linear coefficients and $e_t$ is a Gaussian shock with constant error variance. The OLS estimator $\hat{\bm \alpha_p}$ then provides a linear quantile-specific approximation to the predictive density of the flexible TVP regression. The key advantage of this approach is that it improves interpretability and allows for a straightforward decomposition of the driving forces of right and left-tail forecasts of GDP growth.

Figures \ref{fig:dcomp1q} and \ref{fig:dcomp4q} illustrate the decomposed tail risks in panel A and B. The dashed black line shows the level of tail risks as predicted by our TVP model at the time of the prediction. The bars show the decomposition of the predicted $5^{th}$ and $95^{th}$ percentile approximated by the ten-year rolling linear regression. Panel C and D each show the linear posterior summary's corresponding coefficients to assess the risk indicators' marginal effect. A circle indicates whether a coefficient is significant at the 5\% significance level. Subsequently, we do not discuss the tail risks around WW\rom{2}, because we can not distinguish between the impact of macrofinancial variables and war time effects.

\subsection{Main Results}
\begin{landscape}
\begin{figure}[!htbp]
\includegraphics[width = 1.5\textwidth]{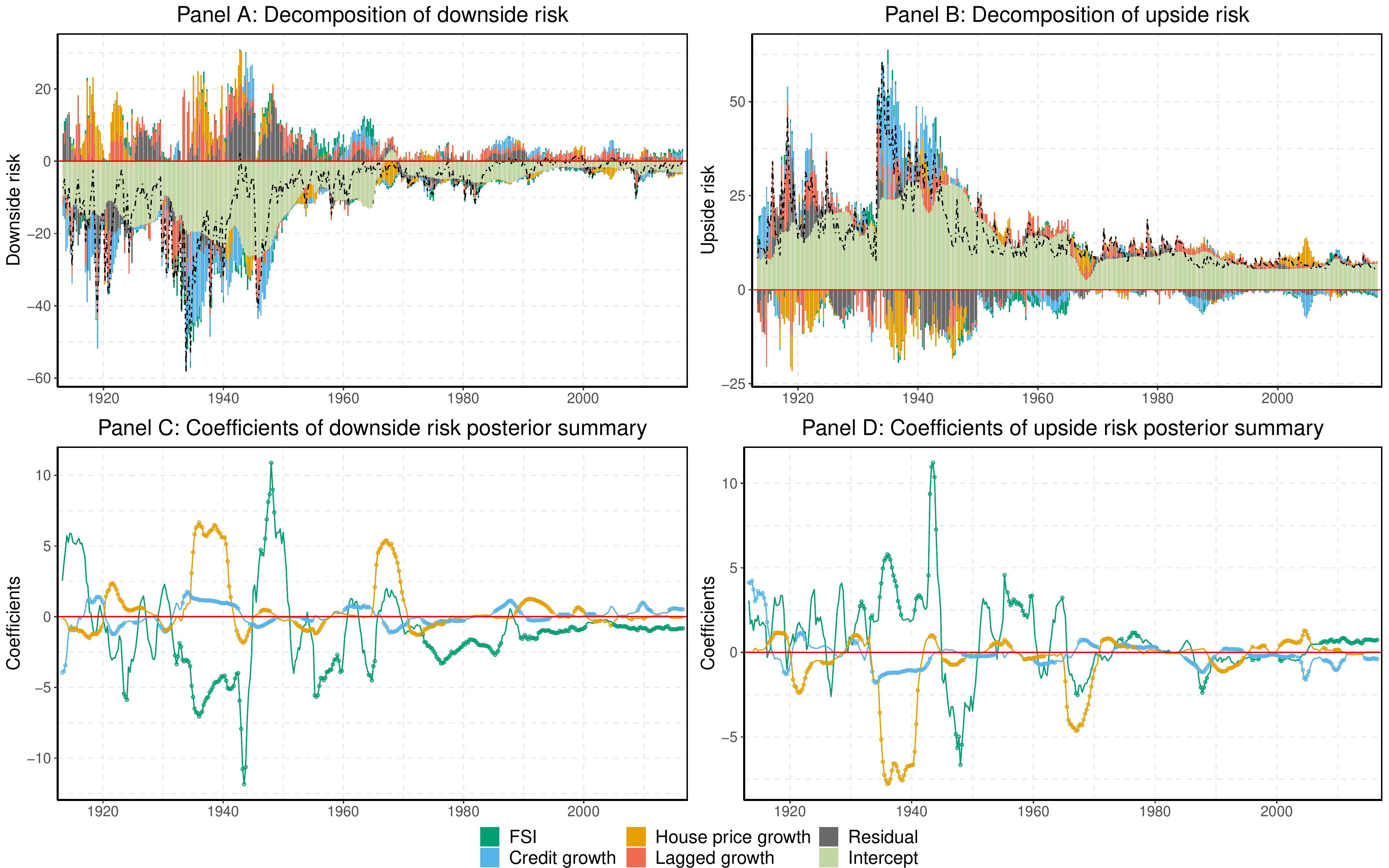}
              \caption{Decomposition of tail risks one quarter ahead}
              \label{fig:dcomp1q}
    \medskip 
    \begin{minipage}{1\columnwidth} 
		{\footnotesize \textit{Note:} Panel A and B show the predictive variable relevance for the predicted $5^{th}$ and $95^{th}$ percentile over time. We approximate the predicted $5^{th}$ and $95^{th}$ percentile using a ten year rolling window linear regression  model. Panel C and D show the coefficients of the linear approximation over time. A circle indicates a significant coefficient (5\% significance level).\par}
	\end{minipage}
\end{figure}
\end{landscape}

\begin{landscape}
\begin{figure}[!htbp]
\includegraphics[width = 1.5\textwidth]{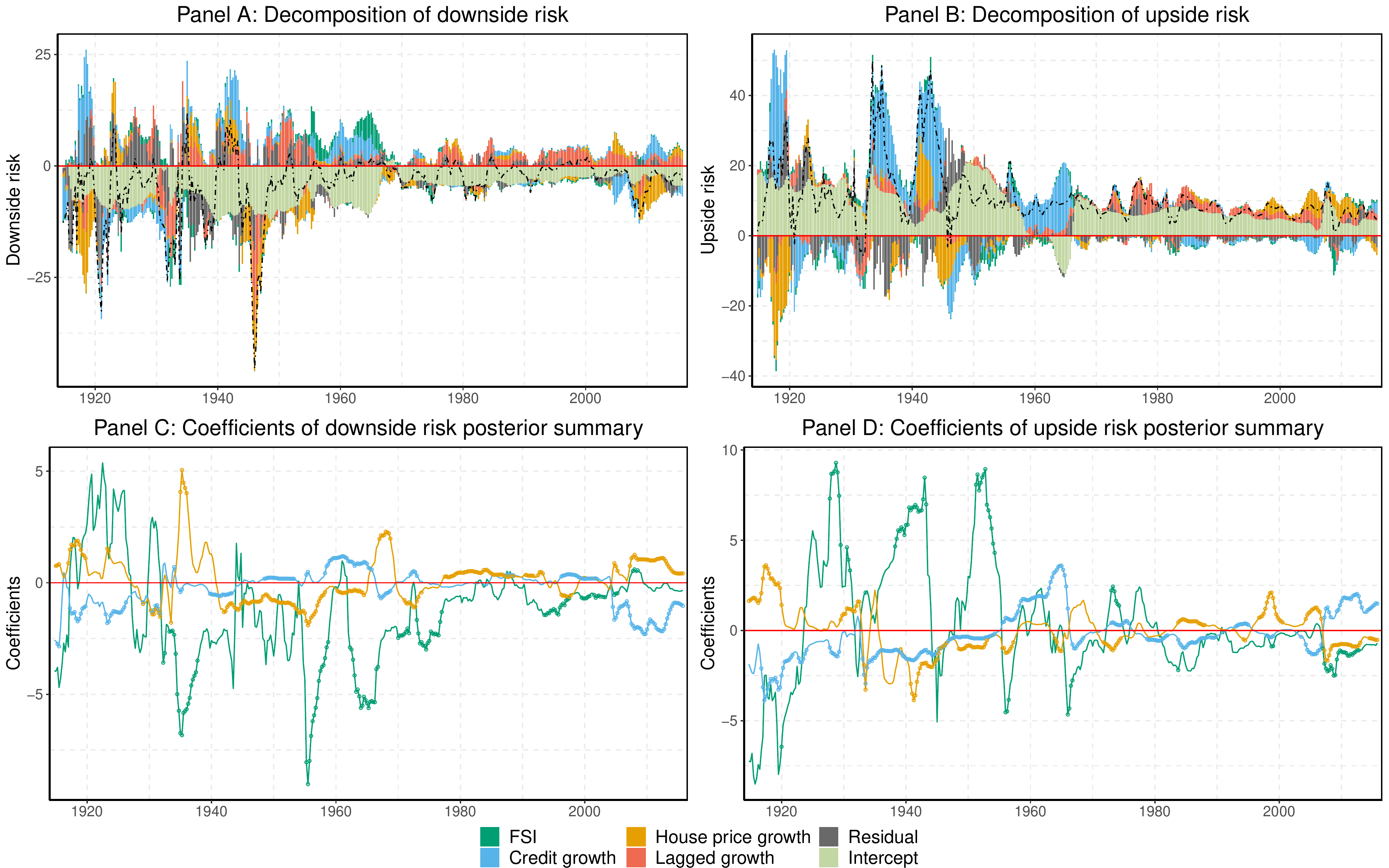}
               \caption{Decomposition of tail risks four quarter ahead}
              \label{fig:dcomp4q}
    \medskip 
    \begin{minipage}{1\columnwidth} 
		{\footnotesize \textit{Note:} Panel A and B show the predictive variable relevance for the predicted $5^{th}$ and $95^{th}$ percentile over time. We approximate the predicted $5^{th}$ and $95^{th}$ percentile using a ten year rolling window linear regression  model. Panel C and D show the coefficients of the linear approximation over time. A circle indicates a significant coefficient (5\% significance level).\par}
	\end{minipage}
\end{figure}
\end{landscape}

Starting with the intercept, we observe a staggering reduction of the intercept post-WW\rom {2} (see Figures \ref{fig:dcomp1q} and \ref{fig:dcomp4q}), which is the conditional average downside risk of the ten-year window of the rolling linear posterior summary. Therefore, the conditional average tail risks nowadays are significantly lower compared to the start of the sample. This result is partly unsurprising since the overall tail volatility has substantially decreased since WW \rom{2}, however the reduction in the conditional average growth risk is still eye-catching.

\cite{adrian2018} find that financial stress is the main short-term factor in predicting left tail risks. This is evident during events such as both oil crises, the dot-com bubble, and the Great Recession, where financial stress greatly contributes to downside risks. However, during the Global Financial Crisis, credit and house price growth were the major contributors to downside risks for the 4-quarter ahead prediction (see Figure \ref{fig:dcomp4q} Panel A). Likewise, before WW\rom{2}, both variables play an important role in predicting growth-at-risk, especially during the early years of the Great Depression. In fact, in midst of the Great Depression, credit growth contributes the most to downside risks. Interestingly, unlike its impact on the left tail, financial stress has little effect on the upper tail of the predicted distribution. Hence, financial stress indeed leads to longer and fatter left tails, but does not influence the upper tail of the predicted distribution. This is reflected in the coefficients, with a significant negative effect on left tail risks, but a less significant effect on upside risks, as seen in Figures \ref{fig:dcomp1q} and \ref{fig:dcomp4q} Panel C and D. More precisely, financial stress has a significant negative marginal effect on left tail risks (panel C), while the marginal effect on upside risks is less significant both economically and statistically.\footnote{Although the OLS coefficient of the one quarter ahead 95$^{th}$ percentile posterior summery is consistently significantly positive since the Global Financial Crisis, the effect is only small.}

Credit growth typically affects downside risks negatively in the medium and long term, but less so at shorter time horizons \citep{adrian2018, galan2020}. While our findings broadly confirm this, we are also able to shed light on temporary and transitional effects over time. In our one-quarter ahead predictions, we find a negative impact of credit growth both after the Global Financial Crisis and the first oil shock, but for completely different reasons. While rapid credit expansion drives downside risks during the first oil crisis, a credit crunch - i.e. too little credit - increased downside risks following the Global Financial Crisis (see Figure \ref{fig:dcomp1q} Panel C). This stands in stark contrast to the effect of credit growth in the four-quarter ahead predictions, where the sharp decline in the coefficient in the mid-2000s results in a large negative overall effect.\footnote{This once again confirms the advantages of our flexible empirical approach. In a similar vein, \citet{aikman2019credit} show, in their 3-year ahead quantile regression GaR estimation, that the sign of the credit growth coefficient depends on the respective sub-sample. For example, using sub-samples until 1997 or 2002, they find a positive effect of credit growth on the 3-year ahead GaR. Only when extending the sub-sample to 2007, they find a negative effect of credit growth. Their -- at first view surprising -- results are therefore consistent with our findings.} Before WW\rom{2}, credit growth was the most important driver of tail risks, contributing to the post-WW\rom{1} recession, the Depression of 1920-1921, as well as the deterioriation in growth risks during the Great Depression. Again, when looking at Panel C in Figures \ref{fig:dcomp1q} and \ref{fig:dcomp4q}, we see the importance of using a flexible approach, as the coefficients vary substantially over time both in terms of sign and magnitude. While financial stress has only a minimal impact on upside risks, credit growth is the  main contributor to higher and lower risks, depending on the respective time period.

The effect of financial stress and credit growth on downside risks to economic growth since the 1970s is well in line with previous literature. For house price growth, we would expect no or only a small effect on downside risks \citep{aikman2019credit,galan2020}; however, we find a significant effect on short-term tail risks in the 1950s, 1960s, early 1990s, and 2000s. Figures \ref{fig:dcomp1q} and \ref{fig:dcomp4q} Panel C offer insights into the patterns observed. Fast-rising house prices during the 1950s, partly due to the GI Bill, and increasing subprime lending in the 2000s negatively impacted economic growth risks by creating a threat of a real estate bubble. Conversely, declining house prices, as seen in the early 1990s and late 1960s, also negatively impacted downside risks, with the most pronounced effect seen in the mid-2000s four-quarter ahead prediction. In this respect, our findings differ from previous studies that suggest little or no effect of house prices on short-term tail risks, as our empirical model is able to capture time-varying effects. Interestingly, house price growth also affects the upside risks of economic growth, which is well in line with concepts of a boom-bust pattern in the financial cycle literature \citep{borio2014}. In the pre-WW\rom{2} period, the effect of house price growth on tail risks was rather small except for the house price crash after WW\rom{1}. Our novel results fit well with the literature about housing and financial cycles, which concludes that over the last century, mortgage lending and total credit to households increased strongly, thereby reinforcing the feedback effects to the whole economy and the impact on tail risks to economic growth. As a result, house prices have become more relevant also from a financial stability perspective, as financial imbalances may have increased with rising leverage in recent decades, compared to the pre-WW\rom{2} period \citep[see, for instance,][]{jorda2015leveraged, jorda2016great, mian2017household}. 

\subsection{Common patterns in times of crises}

To get an in-depth understanding of our empirical results, it is useful to explain the findings by reference to the two major economic crises in our sample. For this purpose, we illustrate the results on the basis of TVP Local Projections during the Great Depression on the one hand, and the Global Financial Crisis on the other. The results show significant fluctuations in coefficients during both crises, with an even more pronounced effect observed during the latter crisis. Interestingly, there are both similarities and differences in the impact of financial risk indicators on downside risks during the two crisis periods, with particularly strong differences in the effect of house prices.

Both during the Great Depression and the Global Financial Crisis, financial stress amplified downside growth risks at short time horizons, in the case of the Great Depression partly also at longer time horizons.\footnote{Please note that the illustrated positive coefficients are not statistically significant, except for 2008Q1, when the positive coefficient four quarters ahead may reflect a partial reversion of the strongly negative effect for the one quarter ahead tail forecast.} In contrast, the effect is somewhat different in the years preceding the crisis, as pre-crisis growth risks are only affected by financial stress prior to the Global Financial Crisis. In both cases, as the crisis unfolded, the effect on longer term horizons disappeared and the impact on short-term (one quarter ahead) predictions intensified. Thus, financial stress elevated downside risks to economic growth in both crises, however, it only influenced pre-crisis growth risks before the Global Financial Crisis.

For credit growth, we observe comparable patterns in both crises, with higher credit growth being associated with heightened growth risks both prior to and during the crisis. During the Great Depression, the coefficients for all horizons shifted upward, suggesting that a credit crunch, rather than excessive credit, heightened growth risks in later years. On the contrary, this upward shift in coefficients was only observed for the one quarter ahead horizons during the Global Financial Crisis, potentially indicating the success of central bank intervention in avoiding a credit crunch. 

While the results are comparable for financial stress and credit growth, the relationship between house price growth and downside risks shows strong differences between the two crises. In the Great Depression, rising house prices had a negative impact on growth risks, whereas during the Global Financial Crisis, falling house prices had the same effect. These results for the Global Financial Crisis are well in line with recent findings on housing wealth effects \citep[e.g.][]{mian2013household}. The literature finds that a decline in house prices can also reduce consumption, especially when housing wealth constitutes a large proportion of households' overall wealth.


\begin{figure}[!htbp]
\includegraphics[width = 1\textwidth]{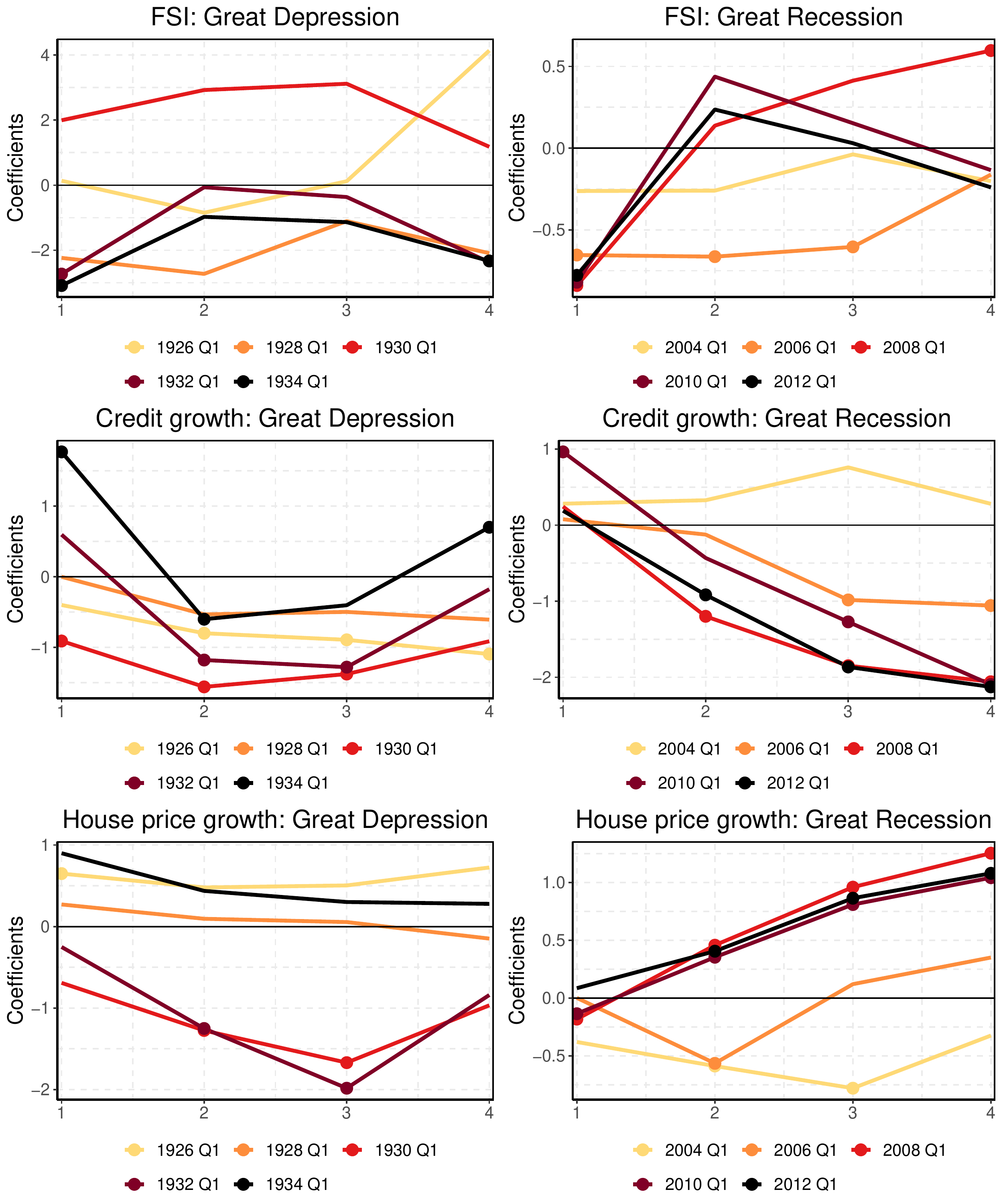}
              \caption{TVP Local Projections}
              \label{fig:pred1q}
    \medskip 
    \begin{minipage}{1\textwidth} 
		{\footnotesize \textit{Note:} This figure shows TVP Local Projections of the linear approximation at different time points. A circle indicates a significant coefficient (5\% significance level).\par}
	\end{minipage}
\end{figure}

\subsection{Discussion}\label{sec: discussion}
In the following, we point out some main findings before discussing more specific effects regarding the financial risk indicators. First, we see that a more flexible approach for estimating GaR is warranted and important, especially when including many indicators (i.e. not only financial stress) and when estimating the effects over a long time period. We find that the link between tail risks and financial risk indicators is time-varying which has strong implications for policy-makers. For instance, when calibrating models for the application of optimal macroprudential policy, such as the one proposed by \citet{suarez2022}, with non time-varying parameters (e.g. quantile regression estimates), this could lead to misleading policy recommendations. Furthermore, not only the effect of risk indicators is time-varying, but also the effect of macroprudential policies can vary over time \citep[see for example][]{jimenez2017, cerutti2017}. Therefore, instead of modeling all kinds of interactions between risk indicators (and possible also macroprudential policies)\footnote{For example whether bubbles are leveraged or not, see e.g. \citet{jorda2015leveraged}.}, it is easier and more effective to use a more flexible empirical approach like the one proposed in this paper. Second, TVP-SV models outperform standard quantile regression models in tail forecasting and therefore give a more accurate picture of future downside risks, which also translates into more reliable warning signals about a potentially sharp recession and a more efficient use of macroprudential polices. 

In a next step, we summarize and discuss the key findings and connect the impact of the individual risk indicators to the broader literature of financial cycles. With respect to financial stress, we find a consistently negative marginal and total effect on downside risks over the whole sample for both one and four quarter ahead predictions. The relationship between financial stress and downside risks to economic growth, as shown by \citet{adrian2019vulnerable}, seems to hold true across time, although with a varying magnitude. Therefore, policy-makers should pay close attention to financial stress and avoid it whenever possible, as it has increased growth risks for over 100 years with no benefit to the upper tail of the distribution. 


The impact of credit growth on tail risks is time-varying and therefore harder to generalize. Once again, this finding confirms that a more flexible approach is warranted when assessing growth risks. For the one quarter ahead predictions, we see  signs of a credit crunch during the Great Depression and the Global Financial Crisis, as indicated by a negative total effect on downside risks, but with a positive marginal effect. This implies that negative credit growth increased the risks to economic growth. However, the stark difference between the two crises is the duration and magnitude of the credit crunch, thus showing the effectiveness of central banks' interventions in 2008 and the following years. Strikingly, while credit growth did not significantly increase downside risks in the second half of the last century, this was the case in the run-up to the Global Financial Crisis, comparable to the Great Depression (both in terms of absolute and marginal effects). These results are in line with the seminal work of  \citet{jorda2015leveraged} who show that credit boom fuelled stock bubbles are much costlier compared to non-credit boom bubbles. Furthermore, \citet{gorton2020good} find that the average credit boom in a sample of 34 countries from 1960 to 2010 last 11 years and is not necessarily bad. Our analysis supports this view, e.g. the credit boom after WW\rom{2} had a very desirable outcome by decreasing downside and increasing upside risks.

Finally, our findings regarding house price growth are also notable. While financial stress has a remarkably consistent impact on growth risks, and the impact of credit growth during the Great Depression and the Global Financial Crisis is also comparable, the effects of house price growth on tail risks during the Great Recession was unprecedented. When house prices collapsed during the Global Financial Crisis, this had a strong impact on downside risks. Interestingly, we see an almost exclusive positive marginal effect of house prices on growth risks since the 1970s. When linking this fact to a rising home ownership ratio, it implies an increasing wealth effect on consumption in the case of collapsing house prices \citep{mian2013household, jorda2017, graham2021house}. This view is also supported by the fact that the correlation between house prices and credit growth increased strongly since the 1960s and remained high thereafter. Therefore, the mortgage-financed housing boom may explain why the effect of house price growth tail risks was unprecedented during the Global Financial Crisis, which is also in line with the broader financial cycle literature \citep{borio2014}.


\section{Conclusion}

Predicting growth risks has become a critical aspect of policymaking since the Global Financial Crisis, and it is particularly important for macroprudential policy-makers. While previous research has identified financial stress, credit growth, and house price growth as key indicators for predicting growth risks, these studies often rely on short time series and inflexible models. In contrast, our research uses a flexible empirical approach and a 130-year historical time series to improve forecasting accuracy and to gain a better understanding with respect to the changing relationships between tail risks and financial risk indicators.

Our findings suggest that certain risk indicators, such as credit growth, have a time-varying effect on tail risks, but tend to follow similar patterns during financial crises. On the other hand, the impact of house prices on tail risks is time-varying for both upside and downside risks, and its impact on growth risks during the Global Financial Crisis was also exceptional compared to historical trends. Additionally, we find that financial stress is consistently associated with increased growth risks (i.e. lower values of GaR) over the entire sample period.

This paper also serves as a note of caution in drawing policy implications from short time series or single historical events, as the observed effects of risk indicators may be time-dependent. In addition to financial risk indicators, policy variables may also have a time-varying impact on tail risks throughout the credit and business cycle. Further research is necessary to assess the effectiveness and importance of macroprudential policy in crisis situations and the potential trade-offs that may occur during normal times.




\newpage

\clearpage
\addcontentsline{toc}{section}{Literatur}
\bibliographystyle{apalike}
\bibliography{ref}

\setcounter{table}{0}
\renewcommand\thetable{A.\arabic{table}}
\setcounter{figure}{0}
\renewcommand\thefigure{A.\arabic{figure}}

\clearpage

\section*{Appendix A: Technical appendix}
We use a Bayesian approach to estimation and inference. This requires choosing suitable shrinkage priors on the coefficients and latent states of the model. To do so, we exploit the non-centered parameterization of the state space model (\citet{fruhwirth2010stochastic}) in Eq. (\ref{eq: tvpsv}):
\begin{equation*}
    y_{t+h} = \bm \beta'_0 \bm x_t + \Tilde{\bm \beta}_t \sqrt{\bm V}_\beta \bm x'_t + \varepsilon_{t+h},
\end{equation*}
with $\bm \beta_0$ denoting a set of $K$ time-invariant coefficeints, $ \Tilde{\bm \beta}_t$ is a vector of normalized states with $j^{th}$ element $\tilde{\beta_{jt}}=(\beta_{jt} - \beta_{j0})/\pm \sqrt{v^2_j}$ and $\sqrt{\bm V}_\beta$ is a diagonal matrix with $\pm \sqrt{v^2_j}$ in the $(j, j)^{th}$ position. 

The state equation of this reparameterized model is given by:
\begin{equation*}
\Tilde{\bm \beta}_t = \Tilde{\bm \beta}_{t-1} + \bm \nu_t, \quad \bm \nu_t \sim \mathcal{N}(\bm 0_K, \bm I_K).
\end{equation*}

The non-centered parameterization can be used to elicit shrinkage priors that allow us to answer the question whether certain elements in $\bm \beta_t$ should be constant or time-varying. This is achieved by using Gaussian priors on $\beta_{j0}$ and $\pm \sqrt{v_j^2}$:
\begin{align*}
    \pm \sqrt{v_j^2}|\tau^2_{j, v} &\sim \mathcal{N}(0, \tau_{j, v}^2), \quad \tau_{j, v}^2|a_v, \lambda_{j, v} \sim \mathcal{G}\left(a_v, \frac{a_v \lambda_{j, v}}{2}\right), \quad \lambda_{j, v}|c_v, \kappa_{v} \sim \mathcal{G}\left(c_v, \frac{c_v}{\kappa_{v}}\right), \\
    \beta_{j0}|\tau^2_{j, \beta} &\sim \mathcal{N}(0, \tau_{j, \beta}^2), \quad \tau_{j, \beta}^2|a_\beta, \lambda_{\beta j} \sim \mathcal{G}\left(a_\beta, \frac{a_\beta \lambda_{\beta, j}}{2}\right), \quad \lambda_{\beta, j}|c_\beta, \kappa_{\beta} \sim \mathcal{G}\left(c_\beta, \frac{c_\beta}{\kappa_{\beta}}\right).
\end{align*}
Here, the parameters $a_v, a_\beta, c_v, c_\beta$ and $\kappa_v, \kappa_\beta$ can be either set by the researcher or additional hyperparameters can be used to infer them from the data. We follow the second approach and specify yet another set of hyperpriors on these parameters. On the rescaled $\lambda_v$ and $\lambda_\beta$ we use F distributed priors and on the remaining hyperparameters we use Beta distributions. The precise prior setting then follows the standard setup discussed in \cite{knaus2021shrinkage} and exercised in the shrinkTVP package.

This prior can be used to discriminate between the following cases. If a given regressor is insignificant for all $t$, the prior shrinks $\beta_{j0}$ and $v_j$ towards zero. If a given regressor is initially unimportant but grows in importance over time, the prior forces $\beta_{j0} \approx 0$ and $v_j > 0$. If a regressor is important and has a time-invariant effect on $y_t$, the prior would imply that $\beta_{j0} \neq 0$ and $v_j \approx 0$. Finally, if a regressor is time-varying and important, the prior would allow for non-zero of $\beta_{j0}$ and $v_j > 0$. All this is achieved automatically through the different scaling parameters.

We simulate from the posterior distribution of the latent states and coefficients using an Markov chain Monte Carlo (MCMC) algorithm. Since all steps are standard (and implemented in the \texttt{R} package \texttt{shrinkTVP}), we only summarize the main steps involved and refer to \cite{knaus2021shrinkage} for more details.

The MCMC algorithm cycles between the following steps:
\begin{itemize}
    \item The latent states are simulated all without a loop from a multivariate Gaussian posterior distribution. This can be achieved efficiently by exploiting sparse algorithms.
    \item The time invariant parameters $\bm \beta_0$ and the diagonal elements of $\sqrt{V}_\beta$ are simulated within a single step from multivariate Gaussian posteriors.
    \item To improve sampling efficiency, an ancillarity-sufficiency interweaving step is introduced that redraws $\bm \beta_0$ from a sequence of Gaussian distributions and the diagonal elements of $\bm V_\beta$ from a generalized inverse Gaussia distribution.
    \item The prior variances and the hyperparameters are simulated either from well known full conditionals detailed in \cite{cadonna2020triple} or by entertaining a Metropolis Hastings update.
    \item The error variances are obtained using the algorithm outlined in \cite{kastner2014ancillarity}.
\end{itemize}

\section*{Appendix B: Further Results}

\subsection*{One quarter ahead}

\begin{figure}[!htbp]
\includegraphics[width = 1\textwidth]{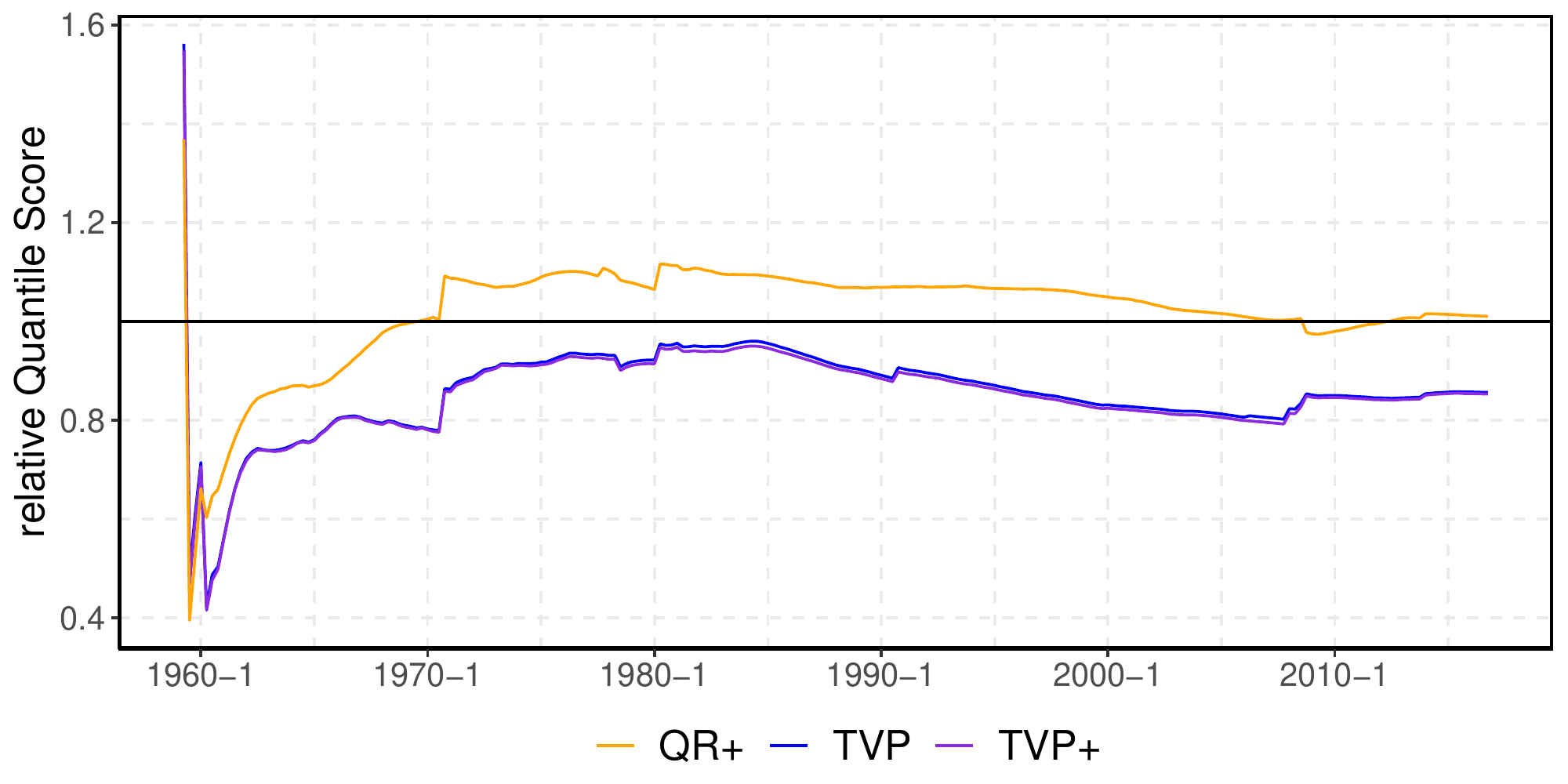}
              \caption{Recursive mean quantile scores (relative to the QR model) one quarter ahead post WW\rom{2}}
              \label{fig:QS1q_post}
    \medskip 
    \begin{minipage}{1\textwidth} 
		{\footnotesize \textit{Note:} This figure shows the cumulative out-of-sample quantile score against the QR model post WW\rom{2}. }
	\end{minipage}
\end{figure}

\begin{figure}[!htbp]
\includegraphics[width = 1\textwidth]{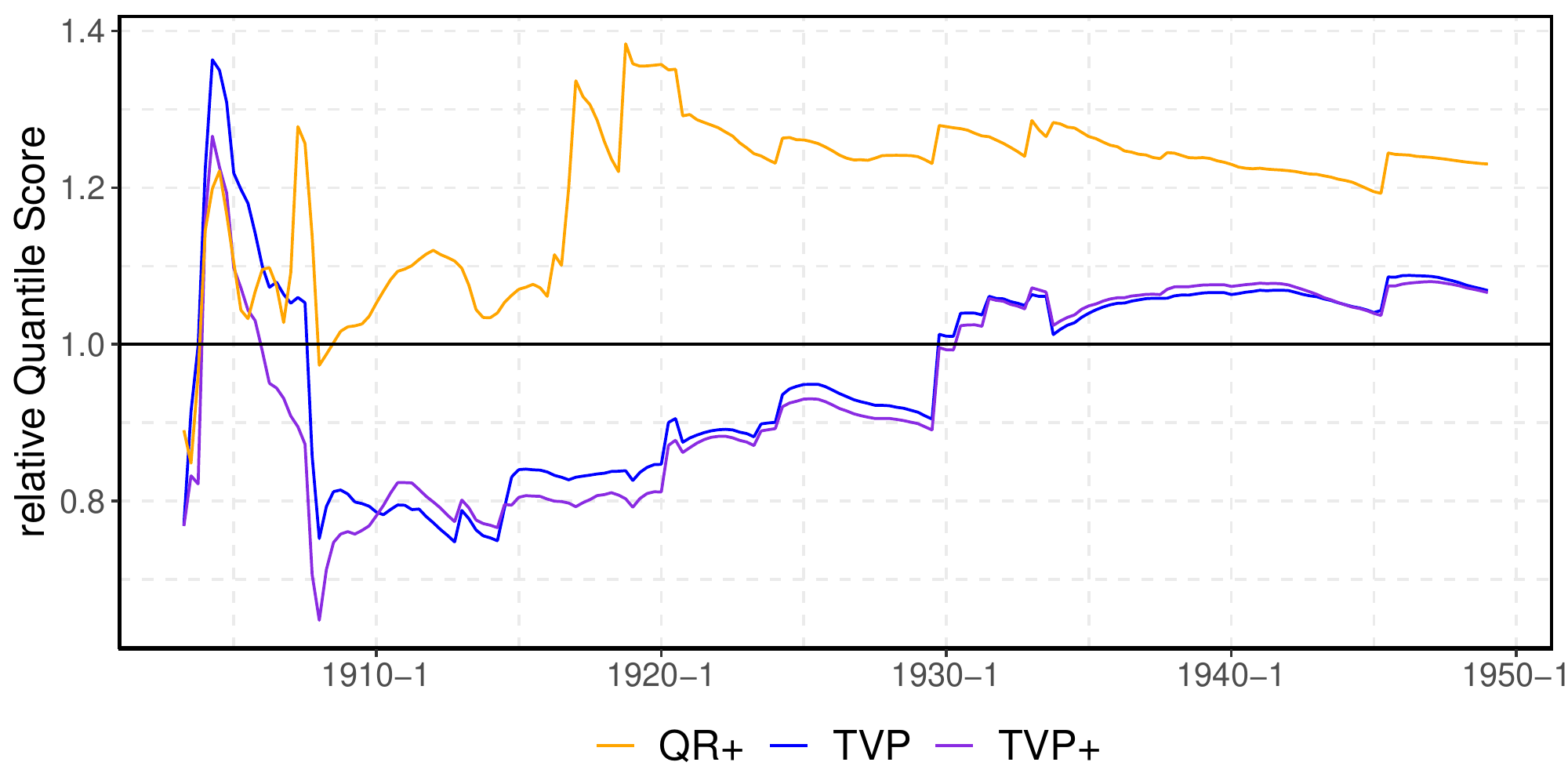}
              \caption{Recursive mean quantile scores (relative to the QR model) one quarter ahead pre WW\rom{2}}
              \label{fig:QS1_pre}
    \medskip 
    \begin{minipage}{1\textwidth} 
		{\footnotesize \textit{Note:} This figure shows the cumulative out-of-sample quantile score against the QR model pre WW\rom{2}. } 
	\end{minipage}
\end{figure}

\begin{figure}[!htbp]
\includegraphics[width = 1\textwidth]{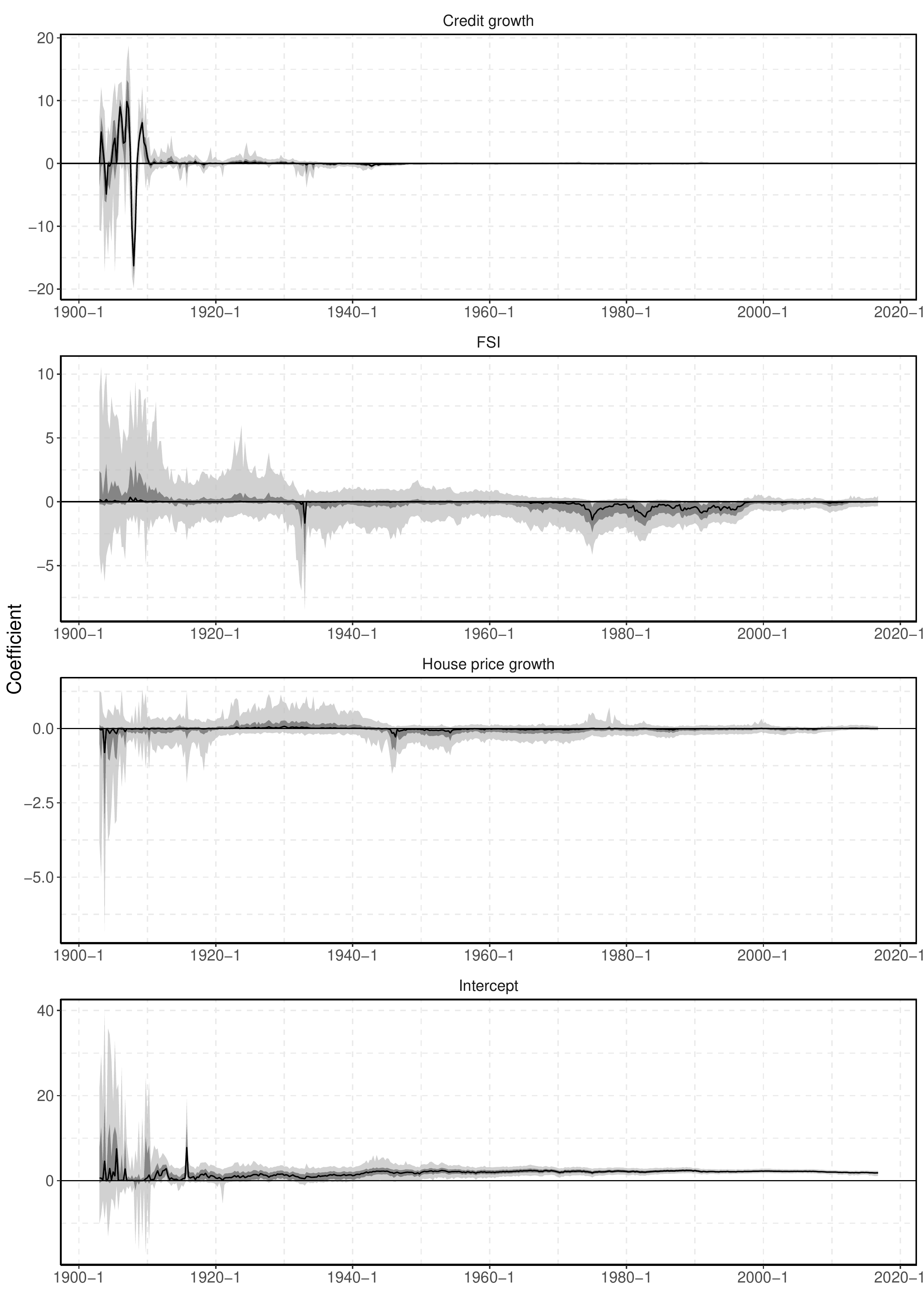}
              \caption{Time series evolution of the coefficients. Recursive estimation one quarters ahead}
              \label{fig:coef1q}
    \medskip 
    \begin{minipage}{1\textwidth} 
		{\footnotesize \textit{Note}: This figure shows the time-varying parameters one quarter ahead. The median is displayed as a black line, and the shaded areas indicate the pointwise 90\% and 50\% posterior credible intervals. \par}
	\end{minipage}
\end{figure}

\newpage
\subsection*{Four quarters ahead}

\begin{figure}[!htbp]
\def\svgwidth{1\textwidth}
        \includegraphics[width = 1\textwidth]{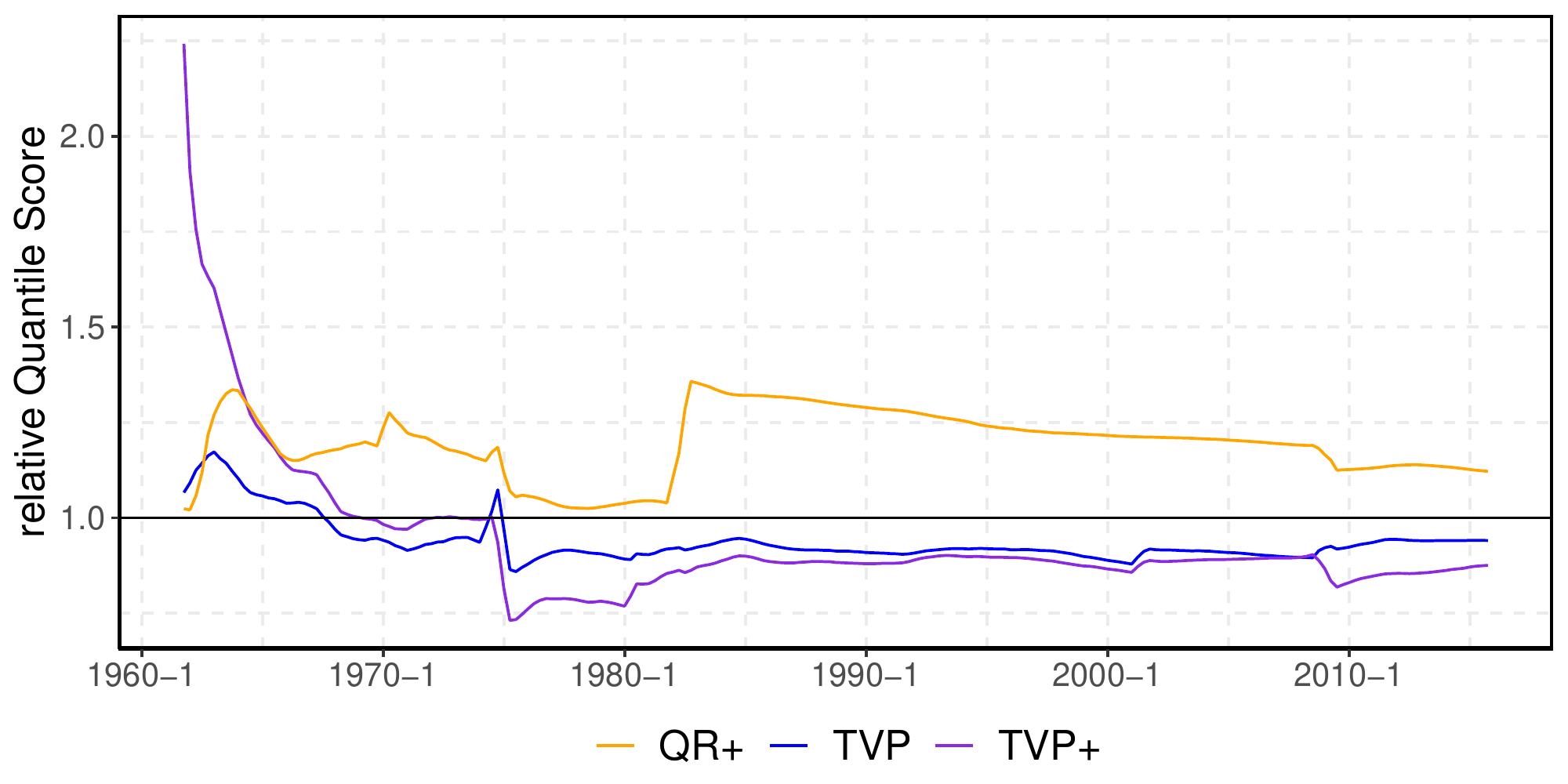}
              \caption{Recursive mean (relative to QR FSI) four quarters ahead post WW\rom{2}}
              \label{fig:QS4q_post}
    \medskip 
    \begin{minipage}{1\textwidth} 
		{\footnotesize \textit{Note:} This figure shows the cumulative out-of-sample Quantile Score against the QR model post WW\rom{2}.}
	\end{minipage}
\end{figure} 

\begin{figure}[!htbp]
\includegraphics[width = 1\textwidth]{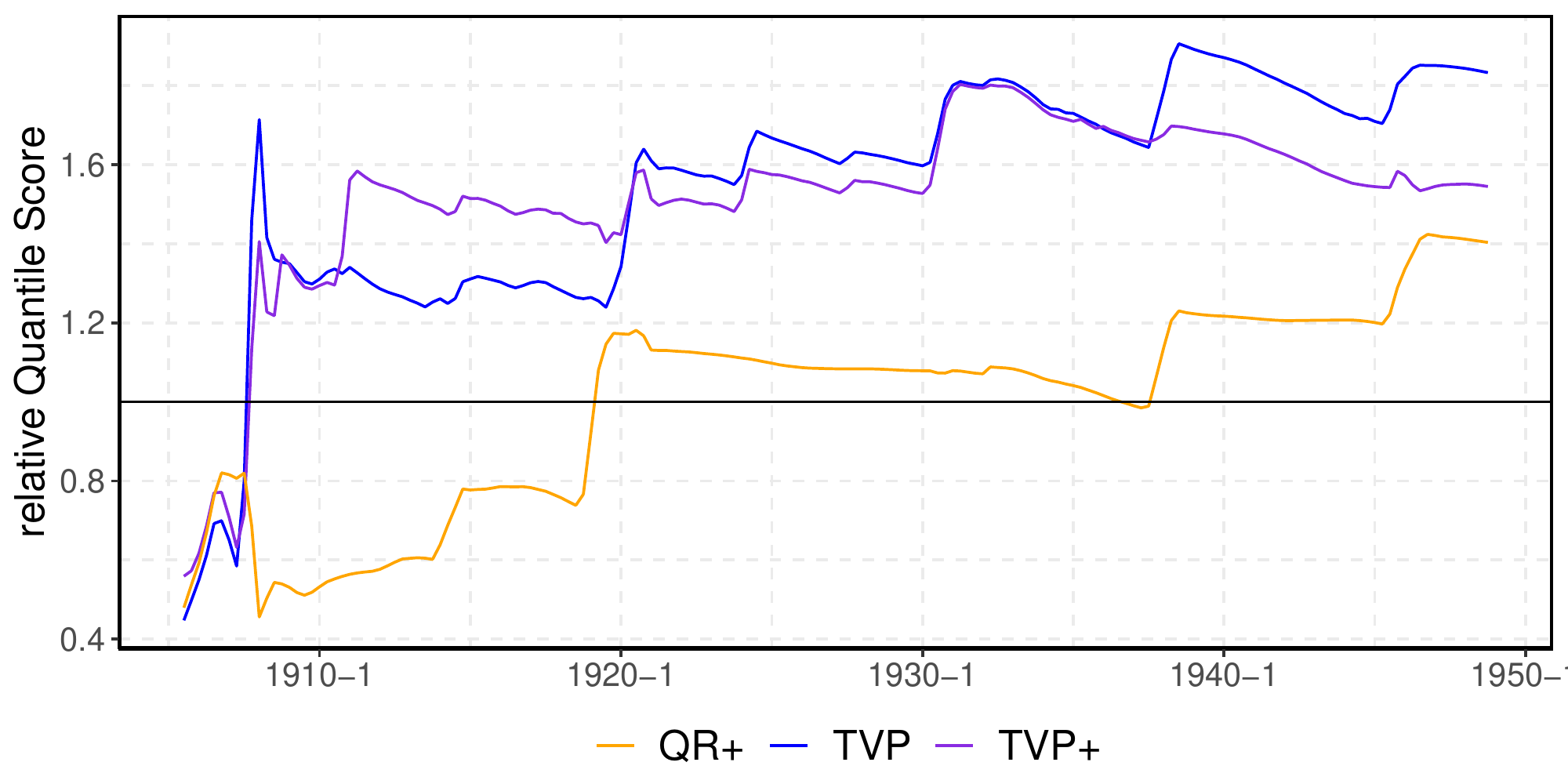}
              \caption{Recursive mean (relative to QR FSI) four quarters ahead pre WW\rom{2}}
              \label{fig:QS4q_pre}
    \medskip 
    \begin{minipage}{1\textwidth} 
		{\footnotesize \textit{Note:} This figure shows the cumulative out-of-sample Quantile Score against the QR model pre WW\rom{2}.}
	\end{minipage}
\end{figure}

\begin{figure}[!htbp]
\def\svgwidth{\columnwidth}
  \includegraphics[width = 1\textwidth]{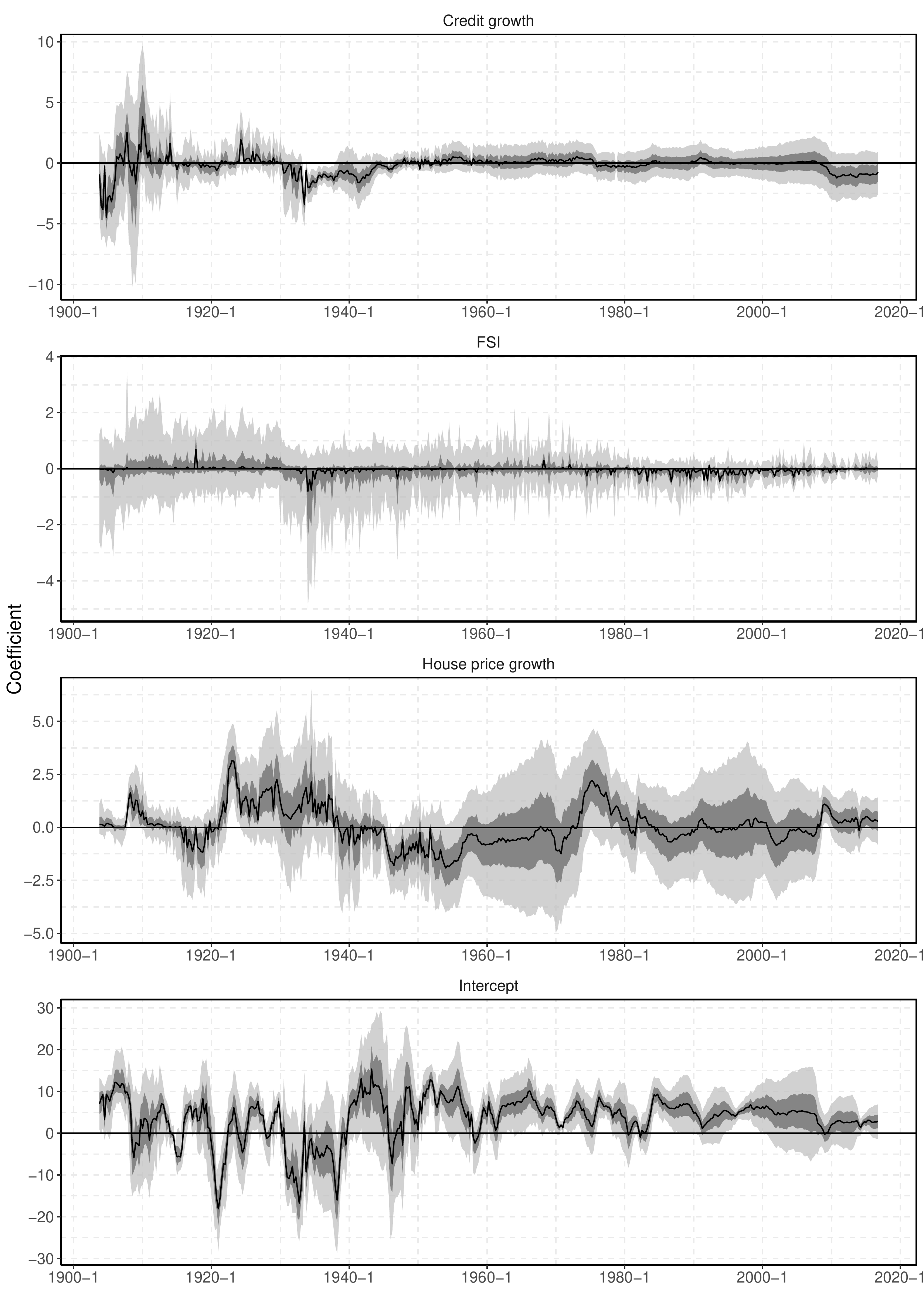}
              \caption{Time series evolution of the coefficients. Recursive estimation four quarters ahead}
              \label{fig:coef4q}
    \medskip 
    \begin{minipage}{1\textwidth} 
		{\footnotesize \textit{Note:} This figure shows the time-varying parameters four quarters ahead. The median is displayed as a black line, and the shaded areas indicate the pointwise 90\% and 50\% posterior credible intervals. \par}
	\end{minipage}
\end{figure}

\end{document}